\documentclass{ifacconf}
\usepackage{mathrsfs} 
\usepackage{graphicx}      
\usepackage{natbib}
\usepackage{amsfonts}
\usepackage{tikz}\usetikzlibrary{arrows.meta}
\usepackage{flushend}
\usepackage{amsmath} 
\usepackage{amssymb}  
\usepackage{enumitem}

\usepackage{algorithm}
\usepackage{algpseudocode}
\usepackage{float}
\usepackage{stfloats}

\DeclareMathOperator*{\argmin}{arg\,min}

\newtheorem{proof}{\textbf{Proof}}
\newtheorem{remark}{\textbf{Remark}}
\newtheorem{theorem}{\textbf{Theorem}}

\newtheorem{definition}{\textbf{Definition}}

\newtheorem{problem}{\textbf{Problem}}

\newcommand{\matrices}[1]{\begin{bmatrix} #1\end{bmatrix}}

\begin{document}
	\begin{frontmatter}
		
		\title{Cryptographic switching functions for multiplicative watermarking in cyber-physical systems \thanksref{footnoteinfo}}
		
		\thanks[footnoteinfo]{
			This work has been partially supported by the Research Council of Norway through the project AIMWind.
		}

		\author{Alexander J. Gallo$^\diamond$, and }\author{Riccardo M. G. Ferrari$^{\diamond}$}\\
		\vspace{.1 in}{\small
			{$^\diamond$\it Delft Center for Systems and Control, Mechanical, Maritime, and Materials Engineering, TU Delft, Delft, Netherlands}\\
			(email: {a.j.gallo,r.ferrari}@tudelft.nl)
		}

		\begin{abstract}                
			In this paper we present a novel switching function for multiplicative watermarking systems.
			The switching function is based on the algebraic structure of elliptic curves over finite fields. 
			The resulting 
			function allows for both watermarking generator and remover to define appropriate system parameters, sharing only limited information, namely a private key.
			Given the definition of the switching function, we prove that the resulting watermarking parameters lead to a stable watermarking scheme.
		\end{abstract}
		
		
	\end{frontmatter}
	
	\section{Introduction}
	Cyber-physical systems (CPS) are a class of systems characterized by high penetration of computational and communication resources, providing increased performance and efficiency.
	Recent events have highlighted that the security of these systems, which aptly describe power grids, transport networks, and many other industrial plants, is of critical importance.
	Indeed, some high profile cases have made it evident that the integration of communication 
	within control systems 
	have also introduced the possibility of malicious agents penetrating these systems and performing attacks \citep{falliere2011w32,lee2008cyber}.
	
	Over the past decade, a growing body of research has been developed in the control systems community to detect, isolate, and mitigate the presence of attackers on control systems \citep{sandberg2015cyberphysical,chen2020guest}.
	Of the possible ways to classify research on cyber-attack detection, one is to distinguish between \textit{passive} and \textit{active} techniques \citep{weerakkody2019resilient}: while in passive methods detection algorithms rely on the effect of the attack and knowledge of the nominal behavior of the system to perform detection, in active architectures specific signals are constructed to enhance the detection capabilities of a diagnostic tool.
	Example of the methods that can be classified as active are additive watermarking \citep{weerakkody2015detecting}, switching multiplicative watermarking \citep{ferrari2020switching}, as well as moving target defense strategies \citep{griffioen2020moving}.
	Here, we focus on developing a design procedure for switching multiplicative watermarking.
	
	The main concept behind the strategy of multiplicative watermarking is to modulate the signals that are to be transmitted over communication networks through two dynamical systems 
	which modulate these signals, increasing the effect of a data injection attack, thus improving detection capabilities.
	Differently to additive watermarking, where a signal is added to the system input to verify its presence in the measured output, multiplicative watermarking does not involve the controlled plant. 
	Rather, modulation of the signal occurs through a watermark generator, before transmission across a communication link. 
	After transmission, the original signal is reconstructed via a watermark remover, before being used for control and diagnostic purposes.
	Thus, if the two systems are appropriately designed, the performance of the control system remains unchanged.
	Introduced in \cite{ferrari2020switching}, switching multiplicative watermarking presupposes that, at certain time instances, the parameters of the watermark generator and remover are changed, synchronously allowing for further detection capabilities against more sophisticated classes of attacks.
	
	
	In this paper, we aim to design a safe switching function, requiring minimal \textit{secret} information to be shared between the watermark generator and remover, thus enhancing its overall security.
	We build a switching function based on modular arithmetic of elliptic curves, which are used in public key exchange cryptography because of the difficulty of solving the discrete logarithm problem on them.
	Our contributions are:
	\begin{enumerate}[label=\alph*.]
		\item we introduce the basic characteristics of elliptic curves, giving a tutorial overview of their properties, and how they may be used for secure control;
		\item we define a switching function for a multiplicative watermarking scheme similar to that proposed in \cite{ferrari2020switching}, providing evidence as to its security properties;
		\item we provide guarantees that the switching function results in a stable watermark generator-remover pair, for finite-impulse response generators.
	\end{enumerate}
	
	
	The remainder of the paper is structured as follows:
	in Section~\ref{ch:probFor} we formally introduce the problem formulation, giving an overview of the modeling of a cyber-physical system, as well as that of the switching multiplicative watermarking detection strategy; 
	in Section~\ref{ch:EC} we give an overview of the mathematical properties of elliptic curves, and specifically of modular arithmetic over elliptic curves;
	in Section~\ref{ch:WM-EC} we present the algorithm defining the switching function, and analyze its properties in terms of safety against attackers and robustness with respect to switching time mismatches.
	Finally, in Section~\ref{ch:Num} we provide a numerical implementation of the algorithm.


	\section{Background and problem formulation}\label{ch:probFor}
\begin{figure}[t]
		\centering
		\includegraphics[width=0.5\linewidth]{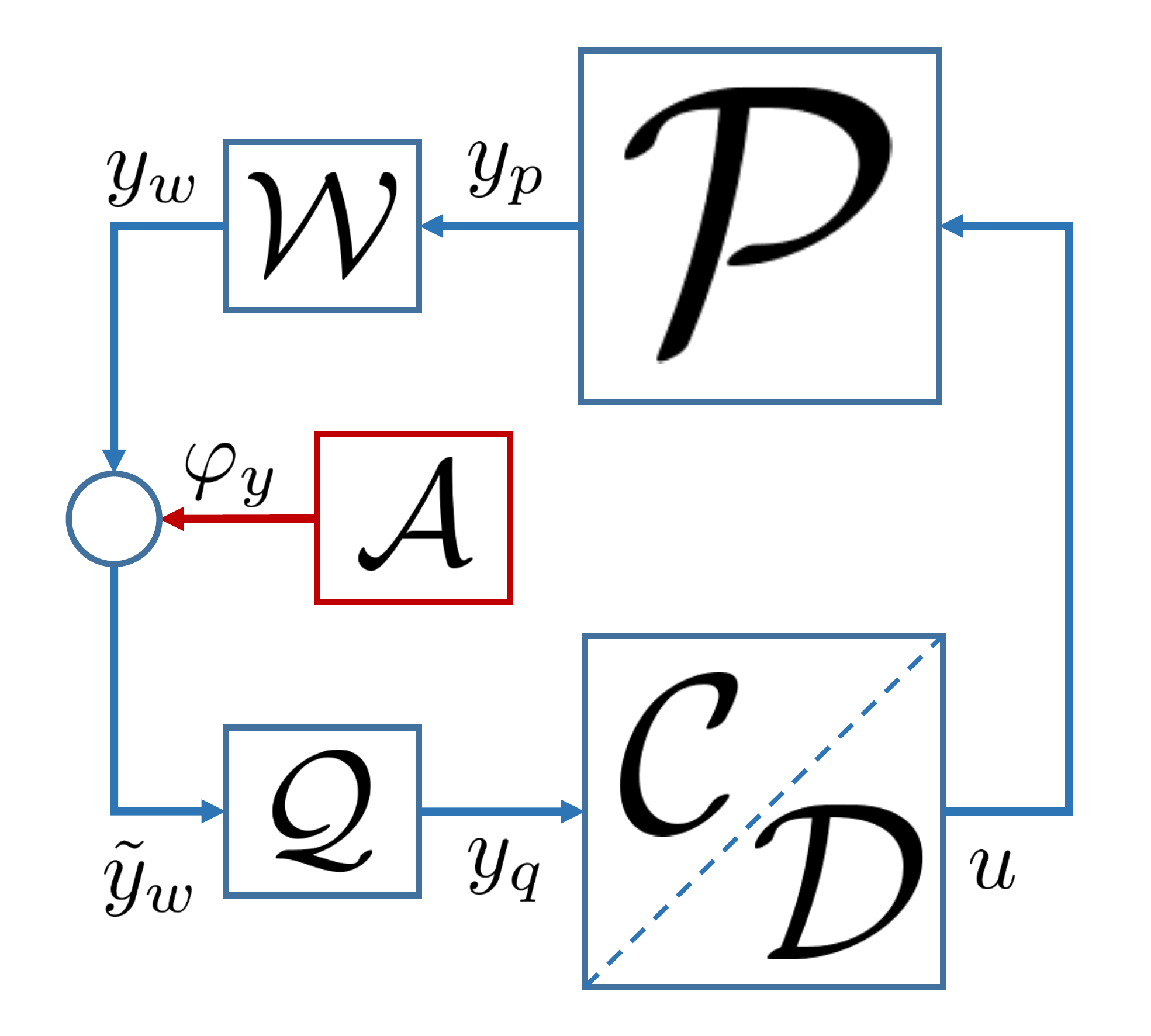}
		\caption{
			Block diagram of the closed loop control system, with plant $\mathcal P$, controller $\mathcal C$, anomaly detector $\mathcal D$, and watermark generator and remover, $\mathcal W$ and $\mathcal Q$. The attacking agent injecting the malicious signal $\varphi_y$ is represented by $\mathcal A$.
		}
		\label{fig:CPS}
	\end{figure}

	In this paper we consider a linear time-invariant (LTI) CPS, represented in Figure~\ref{fig:CPS}.
	The closed loop system is composed of a plant $\mathcal P$ and a controller $\mathcal C$.
	We consider that the output of the plant, $y_p$, is transmitted over some form of communication network to $\mathcal C$, and is therefore exposed to malicious tampering.
	Specifically, we consider a man-in-the-middle attack, capable of eavesdropping the communicated signal, as well as injecting false data into the channel.
	To detect the presence of this attack, we suppose the system is equipped with an anomaly detector $\mathcal D$, as well as a watermark generator and remover pair, $\mathcal W$ and $\mathcal Q$, to enhance the detection properties.
	Specifically, the measurement output $y_p$ is modulated via an LTI system, and the resulting output, $y_w$, is transmitted over the communication network.
	Once received, the signal $\tilde y_w$\footnote{Note here that $\tilde y_w$ is introduced to highlight the possibility of an injected signal in the communication, as will be formally defined in the following.} is demodulated via the watermark remover system $\mathcal Q$, and the resulting output $y_q$ is used as an input to the controller.
	Note that we implicitly assume that the control signal $u$ from the controller to the plant is secure.
	
	In the remainder of this section, we introduce the closed loop dynamics of the CPS, highlighting the properties of $\mathcal W$ and $\mathcal Q$ which make them an appropriate multiplicative watermarking pair, and we formally define the problem.
	
	\subsection{Cyber-physical system}
	We consider the dynamics of the plant $\mathcal P$ to be modeled via the following:
	\begin{equation}\label{eq:dyn}
	\mathcal P : \begin{cases}
	x_p(k+1) = A_p x_p(k) + B_p u(k) + w\\
	y_p(k) = C_p x_p(k) + v
	\end{cases}
	\end{equation}
	where $x_p \in \mathbb{R}^{n_x}$ is the state of the plant, $u \in \mathbb{R}^{n_u}$ is the control input, and $y_p \in \mathbb{R}^{n_y}$ is the measurement output.
	The vectors $w\in\mathbb{R}^{n_x}$ and $v\in\mathbb{R}^{n_y}$ represent the unmodeled disturbances affecting the process and the measurements.
	We suppose that all matrices are of the appropriate dimensions.
	The plant is regulated via a dynamic controller of the form:
	\begin{equation}\label{eq:ctrl}
	\mathcal C : \begin{cases}
	x_c(k+1) = A_c x_c(k) + B_c y_q(k)\\
	u(k) = C_c x_c(k) + D_c y_q(k)
	\end{cases}
	\end{equation}
	where $x_c \in \mathbb{R}^{n_c}$ is the controller state, and $y_q \in \mathbb{R}^{n_y}$ is the output of the watermarking remover system $\mathcal Q$.
	As mentioned previously, $\mathcal Q$ is included in the closed-loop CPS, together with the watermarking generator $\mathcal W$, to enhance the detection capabilities of the anomaly detector $\mathcal D$, to be introduced.
	The watermarking pair is defined by the following dynamical systems:
	\begin{subequations}\label{eq:WM}
		\begin{align}
		&\mathcal W : \begin{cases}\label{eq:WM:W}
		x_w(k+1) = A_w(\theta_w(k)) x_w(k) + B_w(\theta_w(k)) y_p(k)\\
		y_w(k) = C_w(\theta_w(k)) x_w(k) + D_w(\theta_w(k)) y_p(k)
		\end{cases}\\
		&\mathcal Q : \begin{cases}\label{eq:WM:Q}
		x_q(k+1) = A_q(\theta_q(k))x_q(k) + B_q(\theta_q(k)) \tilde y_w(k)\\
		y_q(k) = C_q(\theta_q(k))x_q(k) + D_q(\theta_q(k)) \tilde y_w(k)
		\end{cases}
		\end{align}
	\end{subequations}
	where the vectors $x_w, x_q \in \mathcal R^{n_w}$ and $y_w, y_q \in \mathcal R^{n_y}$ are, respectively, the state and outputs of the watermarking systems $\mathcal W$ and $\mathcal Q$.
	
	Finally, the CPS is equipped with an anomaly detection module $\mathcal D$, to detect the presence of anomalies (such as malicious false-data injection attacks): 
	\begin{equation}
	\mathcal D: \begin{cases}
	x_r(k+1) = A_r x_r(k) + B_r u(k) + K_r y_q(k)\\
	y_r(k) = C_r x_r(k) + L_r y_q(k)
	\end{cases}
	\end{equation}
	where $x_r \in \mathbb{R}^{n_r}$ is the anomaly detector's internal state, and $y_r \in \mathcal R^{n_y}$ its output.
	Note that $y_r$ is used as a residual signal, and the detection test 
	\begin{equation}
	|y_r(k)| \leq \bar{y}_r(k)
	\end{equation}
	is performed at each time instant to detect whether an attack is active on the communication network, where $\bar{y}_r(k)$ is an appropriately defined, time-varying detection threshold, and the inequality is performed component-by-component.
	The definition of $\bar y_p$ is out of the scope of this paper, but may be found in, e.g., \cite{zhang2002robust}.

	\subsection{Attack model}
	As mentioned previously, we consider the output-side signal $y_w$ to be transmitted over a communication network. Thus it is possibly subject to attacks from some time $k_a > 0$.
	We define the received signal, appearing in \eqref{eq:WM:Q}, as:
	\begin{equation}
	\tilde y_w(k) = y_w(k) + \beta_a(k-k_a)\varphi_y(Y_{w,[k-N_a:k]},k),
	\end{equation}
	where $\varphi_y(\cdot)$ is the maliciously defined attack signal, depending on the matrix:
	\begin{equation}
	Y_{w,[k-N_a:k]} \doteq \matrices{y_w(k-N_a) &y_w(k-N_a+1) &\dots &y_w(k)},
	\end{equation}
	and $\beta_a(\cdot)$ is an activation function, which we here consider to be $\beta(\kappa) = 0, \forall \kappa < 0$, and $\beta(\kappa) = 1$ otherwise.

	\subsection{Multiplicative watermarking: some background}
	Let us now focus on the design of the time-varying systems $\mathcal W$ and $\mathcal Q$, i.e., the watermarking pair.
	The results reported in this section rely on those in \cite{ferrari2020switching}.
	\begin{definition}\label{def:WM}
		For two systems $\mathcal W$ and $\mathcal Q$, defined in \eqref{eq:WM}, to be an appropriate watermarking pair, the following conditions must be satisfied:
		\begin{enumerate}[label=\alph*.]
			\item $\mathcal W$ is stable and invertible;
			\item $\mathcal Q$ is stable;
			\item if $\theta_w = \theta_q$, $\mathcal Q$ is the inverse of $\mathcal W$.
			$\hfill\triangleleft$
		\end{enumerate}
	\end{definition}
	
	The definition of the systems $\mathcal W$ and $\mathcal Q$ in \eqref{eq:WM} are parametrized by the vectors $\theta_w,\theta_q \in \mathbb{R}^{n_\theta}$ defining the dynamics of the system.
	These parameters are piecewise constant, and are updated only at specific switching times, to be defined later, with the updates given by:
	\begin{subequations}\label{eq:WM:sw}
		\begin{align}
		&\mathcal W : \begin{cases}
		\theta_w^+(k) = \sigma_w(\mathcal{I}_w(k)), \hfill\text{if }\tau_w(y_p(k)) = 1\\
		x_w^+(k) = \rho_w(x_w^-(k),y_p(k),\theta_w^-(k),\theta_w^+(k))
		\end{cases}\\
		&\mathcal Q : \begin{cases}
		\theta_q^+(k) = \sigma_q(\mathcal I_q(k)), \hfill\text{if }\tau_q(\tilde y_w(k)) = 1\\
		x_q^+(k) = \rho_q(x_q^-(k), \tilde y_w^+(k), \theta_q^-(k), \theta_q^+(k))
		\end{cases}
		\end{align}
	\end{subequations}
	where $\mathcal{I}_w(k), \mathcal{I}_q(k)$ are the sets of information available to $\mathcal W$ and $\mathcal Q$ at time $k$, as defined in Definitions~\ref{def:WM:info:W} and \ref{def:WM:info:Q};
	$\rho_i : \mathbb{R}^{n_w}\times \mathbb{R}^{n_y} \times \mathbb{R}^{n_\theta}\times \mathbb{R}^{n_\theta} \rightarrow \mathbb{R}^{n_w}, i \in \{w,q\}$ is a \textit{jump map} of $\mathcal W$ and $\mathcal Q$,
	$\sigma_i :\mathbb{R}^{n_y} \rightarrow \mathbb{R}^{n_\theta}, i \in \{w,q\}$ is a \textit{switching map},
	and the superscripts $+$ and $-$ denote the values of the vectors before and after a switch.  
	Finally, $\tau_i, i \in \{w,q\}$ are \textit{triggering functions} inducing the switch. 
	
	\begin{definition}
		The functions $\tau_q, \tau_w \in \mathbb{R}^{n_y} \rightarrow \{0,1\}$ are said to be trigger functions of $\mathcal W$ and $\mathcal Q$ if the \textit{triggering sets} $\mathcal C_w \doteq \{y_p : \tau_w(y_p) = 1\}$ and $\mathcal C_q \doteq \{\tilde y_w: \tau_q(\tilde y_w) = 1\}$ are convex and open. The sequences $\mathcal{K}_w \doteq \{k: \tau_w(y_p(k)) = 1\}$ and $\mathcal K_q \doteq \{k: \tau_q(\tilde y_w(k)) = 1\}$ are called the \textit{switching time sequences} of $\mathcal W$ and $\mathcal Q$.
		$\hfill\triangleleft$ 
	\end{definition}
	
	Given that $\mathcal W$ and $\mathcal Q$ are not colocated, it is important to properly define what information is available to $\mathcal W$ and $\mathcal Q$.
	We therefore formally introduce the \textit{information sets}, $\mathcal I_w$ and $\mathcal I_q$, defined in terms of input and output data over a window of size $N_I \geq 1, N_I \in \mathbb{Z}_+$, as well as $\theta_i, i \in \{w,q\}$.
	\begin{definition}[Information at $\mathcal W$]\label{def:WM:info:W}
		The information available to the watermark generator $\mathcal W$ is defined as:
		\begin{equation}
		\mathcal I_w(k) = \{Y_{p,k-N_I,k}, x_w(k-N_I),\theta_w(k)\}.
		\end{equation}
		$\hfill\triangleleft$
	\end{definition}
	\begin{definition}[Information at $\mathcal Q$]\label{def:WM:info:Q}
		The information available to the watermark remover $\mathcal Q$ is defined as:
		\begin{equation}
		\mathcal I_q(k) = \{\tilde{Y}_{w,k-N_I,k}, x_q(k-N_I),\theta_q(k)\};
		\end{equation}
		in addition, for $\kappa_w \in \mathcal K_w$, $\mathcal I_q^+(\kappa_w) = I_q(\kappa_w) \cup \{y_w^+(\kappa_w)\}$.
		$\hfill\triangleleft$
	\end{definition}
	\begin{remark}
		Given the definition of the information sets, $\mathcal I_q \subset \mathcal I_w$ holds, in nominal conditions, for all $k$.
		$\hfill\triangleleft$
	\end{remark}

	\subsection{Synchronized switching}
	To guarantee that the multiplicative watermarking scheme does not influence the closed-loop performance of the system, the following must hold:
	\begin{enumerate}[label=\alph*.]
		\item $\mathcal K_w = \mathcal K_q$ (\textit{synchronized switching times});
		\item the output of $\sigma_w(\kappa) = \sigma_q(\kappa)$ and $\rho_w(\kappa) = \rho_q(\kappa)$, for all $\kappa \in \mathcal K_w$ (\textit{synchronized switches and jumps});
		\item $y_q^+(\kappa) = y_p(\kappa)$ (\textit{synchronized output}). 
	\end{enumerate}
	
	Here we have slightly abused notation, writing $\sigma_w(k)=\sigma_q(k)$, rather than $\sigma_w(\mathcal{I}_w(k)) = \sigma_q(\mathcal{I}_q(k))$.
	We consider the same \textit{induced synchronization} scheme presented in \cite{ferrari2020switching}, by which $\mathcal W$ initializes the switch triggering and $\mathcal Q$ updates its parameters, without any additional communicated data.
	
	\begin{definition}[Synchronized watermarking]
		The watermarking generator $\mathcal W$ and remover $\mathcal Q$ are said to be synchronized if at switching time $k$ they are:
		\begin{enumerate}[label=\alph*.]
			\item trigger synchronized, i.e. $\tau_w(y_p(k)) = \tau_q(\tilde y_w(k)) = 1$;
			\item switch synchronized, i.e. $\theta_w^+(k) = \theta_q^+(k)$;
			\item jump synchronized, i.e. $x_w^+(k) = x_q^+(k)$;
			\item output synchronized, i.e. $y_q^+(k) = y_p(k)$.
			$\hfill\triangleleft$ 
		\end{enumerate}
	\end{definition}
	
	\begin{remark}
		Under synchronized watermarking $\mathcal K_w = \mathcal K_q$ holds, and therefore $\mathcal I_q^+(\kappa_w) \subset \mathcal I_w(\kappa_w), \forall \kappa_w \in \mathcal K_w$.
		$\hfill\triangleleft$
	\end{remark}

	\subsection{Problem formulation}
	While in \cite{ferrari2020switching} detailed techniques are presented to define $\rho_i(\cdot)$, $i \in \{w,q\}$, as well as a minimally \textit{visible}\footnote{Interested readers are referred to \cite{ferrari2020switching} for a definition of switch visibility.} definition of $y_w^+(\kappa_w), \kappa_w \in \mathcal K_w$, 
	the definition of the switching maps $\sigma_i, i \in \{w,q\}$ are left unspecified.
	Thus, the objective of this paper is as follows:
	\begin{problem}\label{prob:Switching}
		Given a switching multiplicative watermarking scheme defined by \eqref{eq:WM}-\eqref{eq:WM:sw}, define switching functions $\sigma_w(\mathcal I_w(k))$ and $\sigma_q(\mathcal I_q(k))$ such that:
		\begin{enumerate}[label=\alph*.]
			\item the entire sequence must not be known a priori;
			\item $\theta_w^+(k) = \theta_q^+(k)$, for all $k \in \mathcal K$;
			\item $\theta_i^+(k)$ is such that $\mathcal W$ and $\mathcal Q$ satisfy the conditions in Definition~\ref{def:WM}.
			$\hfill\triangle$
		\end{enumerate}
	\end{problem}
	
	In the following section, we give some background on the mathematics of elliptic curves, and how they have been used in cryptography to generate private keys for encryption in the IT-security literature.
	After this, in Section~\ref{ch:WM-EC}, we present the algorithm used to generate the parameters of $\mathcal W$ and $\mathcal Q$, highlighting the required information.
	
	\section{Elliptic curves on finite fields: some background}\label{ch:EC}
	Let us now present some background on the arithmetic of elliptic curves.
	We here rely on overviews presented in \cite{wohlwend2016elliptic,lopez2000overview}.
	Elliptic curves are abelian varieties, which have had large success in the field of cryptography.
	Indeed, as outlined in \cite{wohlwend2016elliptic,lopez2000overview}, elliptic curve cryptography (ECC), 
	is a form of \textit{asymmetric} or \textit{public key} cryptography, which guarantees higher levels of security than the Diffie-Hellman-Markle key exchange, or RSA.
	We exploit the mathematical properties of elliptic curves to define a switching function common to $\mathcal W$ and $\mathcal Q$ which, although still including a shared secret, does not require the entire switching sequence to be defined \textit{a priori}.
	
	\subsection{Some fundamentals in group theory}\label{ch:EC:gT}
	To properly introduce both the group defined by the elliptic curve and its operations, let us introduce some definitions.
	
	\begin{definition}
		A \textit{group} is defined as a set $G$ together with a binary operation $\circ$ closed in $G$, i.e., for any $a,b \in G$, $a \circ b \in G$, such that the following axioms (the so called \textit{group axioms}) hold:
		\begin{enumerate}[label=\alph*.]
			\item the operation $\circ$ is associative, i.e., $(a\circ b)\circ c = a\circ (b\circ c)$;
			\item there exists an identity element in $G$, $e$, such that $a \circ e = e \circ a = a$;
			\item there exists an inverse element of $a \in G$, denoted $a^{-1}$, under the group operation, such that $a\circ b = b \circ a = e$.
		\end{enumerate}
		$\hfill\triangleleft$
	\end{definition}
	
	\begin{definition}
		A group $G$ is \textit{abelian} if its binary operator $\circ$ is commutative, i.e. $a\circ b = b \circ a, \forall a, b \in G$.
		$\hfill\triangleleft$
	\end{definition}
	
	\begin{definition}
		A group $G$ is said to be cyclic if, for an element $h \in G$, every element $g \in G$ satisfies $g = xh$ or $g = h^x$, $x\in\mathbb{Z}_+$, depending on whether the group operation is additive or multiplicative.
		The element $h$ is said to be the generator of the group.
		$\hfill\triangleleft$
	\end{definition}
	
	\begin{definition}
		A set $H \subseteq G$ is said to be a cyclic subgroup if it is cyclic, given some generator $h$.
		$\hfill\triangleleft$
	\end{definition}
	
	\begin{definition}
		Suppose $a \in G$, and $e$ is the identity element. The \textit{order} of $a$ is the smallest integer $n$ such that:
		\begin{equation}
		\underbrace{a\circ a\circ \dots \circ a}_n = e.
		\end{equation}
		The set $\{a, a^2, \dots, a^n\}$ (or $\{a,2a,\dots,na\}$) forms a cyclic subgroup of $G$ of order $n$, with $a$ as its \textit{generator}.  
		$\hfill\triangleleft$
	\end{definition}

	\begin{definition}
		A \textit{field} is an algebraic structure composed of a set $\mathbb{F}$ together with the binary addition and multiplication operations, $+$ and $\times$, satisfying the following:
		\begin{enumerate}[label=\alph*.]
			\item $\mathbb{F}$ is an abelian group under addition $+$;
			\item $\mathbb{F} \backslash \{0\}$ is an abelian group under multiplication $\times$;
			\item $\times$ is distributive over addition: $a\times (b+c) = a\times b + a \times c$.
		\end{enumerate}
		A field is said to be \textit{finite} if $|\mathbb{F}|< \infty$.
		$\hfill\triangleleft$
	\end{definition}
	Examples of fields are the set of rational, real, complex numbers $\mathbb{Q}, \mathbb{R}, \mathbb{C}$.
	A field that is fundamental in cryptography, and used in this paper, is the set of integers modulo $l$, $\mathbb{Z}/l\mathbb{Z}$, with $l$ prime. We use the notation $\mathbb{F}_l = \mathbb{Z}/l\mathbb{Z}$ for compactness; this must not be confused with the set of $l$-adic integers.
	
	\subsection{Elliptic curves}
\begin{figure}[t]
		\centering
		\includegraphics[width=0.9\linewidth]{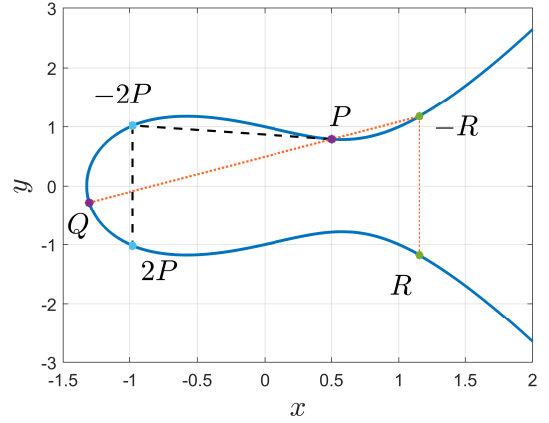}
		\caption{
			The elliptic curve defined in $\mathbb{R}$, with parameters $a = -1$ and $b = 1$, with geometric representation of the operations $R = P + Q$ and $2P$.
		}
		\label{fig:EC}
	\end{figure}

	Given the preliminary definitions in Section~\ref{ch:EC:gT}, we can now introduce elliptic curves.
	Although we are interested in elliptic curves over finite fields, as is further highlighted in the following, we start by presenting the general concept of an elliptic curve over a (possibly infinite) field $\mathbb{F}$:
	an elliptic curve $E(\mathbb{F})$ is the set of points in $\mathbb{F}$, satisfying
	\begin{equation}
	y^2 + a_1 x y + a_2 y = x^3 + a_3 x^2 + a_4 x + a_5.
	\end{equation}
	There are a number of special elliptic curves, of which the Weierstrass normal form:
	\begin{equation}\label{eq:EC:Wierstrass}
	y^2 = x^3 + ax + b,
	\end{equation}
	with $a, b \in \mathbb{F}$, is commonly used in cryptography.
	In Figure~\ref{fig:EC} we present an example of an elliptic curve defined for $\mathbb{F} = \mathbb{R}$, with $a = 1, b = -1$.
	
	The curve $E(\mathbb{F})$ forms an abelian group together with an addition operator $+$, with the so called point at infinity $O$ as its identity.
	Specifically, addition in $E(\mathbb{F})$ satisfies: 
	\begin{itemize}
		\item[--] $P + O = O + P = P$;
		\item[--] for $P, Q \in E(\mathbb{F})$, $P \neq \pm Q$, the point $R = P+Q$ is the point satisfying $P + Q - R = O$;
		\item[--] if $P = (x,y) \in E(\mathbb{F})$, then the negative of $P$, $-P = (x,-y)$, is such that $P - P = O$.
	\end{itemize}
	Coordinates $(x_R,y_R) = R = P+Q$ can be computed via: 
	\begin{equation}
	\begin{array}{cc}
	\left\{
	\begin{array}{l}
	x_R = \lambda^2 - x_P - x_Q \\
	y_R = \lambda(x_P - x_R) - y_P
	\end{array}
	\right.,
	&\lambda = \frac{y_Q - y_P}{x_Q - x_P}
	\end{array}
	\end{equation}
	with $P = (x_P,y_P)$ and $Q = (x_Q,y_Q)$. This can be interpreted geometrically by taking the line through $P$ and $Q$ and finding where it intersects the elliptic curve: this is $-R$. Thus, taking the inverse of the $y$ coordinate, $R = P+Q$ is found. 
	
	Given operator $+$, we are also interested in the so called \textit{doubling operator}, i.e., computing $2P = P+P$.
	Clearly, in this case the geometric interpretation we provided for addition over the elliptic curve does not hold.
	Instead, $2P$ can be interpreted geometrically by taking the tangent to $P$, which crosses the elliptic curve in one point, $-2P$. Then point $2P$ is found by inverting the $y$ coordinate. Geometric representations for addition and doubling are given in Figure~\ref{fig:EC}.
	In coordinate terms, $(x_{2P},y_{2P}) = 2P$ can be found through:
	\begin{equation}
	\begin{array}{cc}
	\left\{
	\begin{array}{l}
	x_{2P} = \lambda^2 - 2x_P\\
	y_{2P} = \lambda(x_P - x_{2P}) - y_P
	\end{array}
	\right.,
	&\lambda = \frac{3x_P^2 + a}{2y_P}
	\end{array}.
	\end{equation}
	
	We can also define \textit{scalar multiplication} over $E(\mathbb{F})$, by repeatedly adding a point $P \in E(\mathbb{F})$ to itself:
	\begin{equation}
	sP = \underbrace{P + P + \dots + P}_{s},
	\end{equation}
	with $s \in \mathbb{Z}_+$.
	For elliptic curves over finite fields, to be used in the definition of the switching function in the following, this operation can be efficiently computed in $O(\log (s))$ using the \textit{Double and Add} algorithm \citep{wohlwend2016elliptic}.
	Scalar multiplication can also be used to define a cyclic subgroup with generator $P$ and order $n$.
	
	\begin{definition}
		The order of a point $P \in E(\mathbb{F})$ is defined as the minimum $n \in \mathbb{Z}_+$ such that:
		\begin{equation}
		\underbrace{P + \dots + P}_{n} = O.
		\end{equation}
		$\hfill\triangleleft$
	\end{definition}
	\begin{definition}\label{def:EC:cof}
		The cofactor of a point $P \in E(\mathbb{F}_p)$ of order $n$ is defined as:
		\begin{equation}
		h \doteq \frac{|E(\mathbb{F})|}{n}.
		\end{equation}
		$\hfill\triangleleft$
	\end{definition}
	
	Lagrange's theorem states that the order of a subgroup must be a divisor of the order of the group: thus, $h\in\mathbb{Z}_+$, for all $P \in E(\mathbb{F}_p)$ \citep{birkhoff2017survey}.
	
	Until now we have described elliptic curves on a generic field $\mathbb{F}$.
	However, for the purpose of our contribution, we are interested in elliptic curves defined on finite fields, specifically $\mathbb{F}_p$, the field of integers modulo $p$.
	These curves are different to that shown in Figure~\ref{fig:EC}, and an example is found in Figure~\ref{fig:EC_fin}.
	This field is used also in Elliptic Curve Cryptography (ECC), a successful public key cryptographic scheme, summarized briefly in Section~\ref{ch:EC:ECC}.

	\subsection{Elliptic curve cryptography}\label{ch:EC:ECC}
	One of the main applications that elliptic curves have found is that of elliptic curve cryptography (ECC).
	Similar to Diffie-Hellman-Merkle private key cryptography, it is based on the difficulty of solving the discrete logarithm problem, i.e., even if $P$ and $S = lP$ are known, there are no efficient solutions to find $l$.
	
	The Diffie-Hellman key exchange on an elliptic curve has the following structure: 
	suppose Alice and Bob want to share a key, without any shared secrets. 
	Additionally, suppose they agree on the parameters of a shared (public) elliptic curve $E(\mathbb F_p) = \{\mathbb F_p, P, n, h, a, b\}$, where $\mathbb{F}_p$ is the field on which the elliptic curve is defined, $P$ is a generator point, $n$ and $h$ are respectively the order and cofactor of $P$, and $a,b$ are the parameters of the elliptic curve  \eqref{eq:EC:Wierstrass}.
	Alice defines a private key $k_a$, and computes $Q_a = k_aP$, her public key. 
	Similarly, Bob defines a private key $k_b$, and generates public key $Q_b = k_bP$.
	Thus, Alice and Bob exchange their public keys, and compute the shared key, $Q_{ab} = k_aQ_b = k_b Q_b = (k_a k_b)P$.		
	This shared information is reached exploiting private information that has not been transmitted over a (possibly insecure) communication channel.
	Any malicious agent capable of eavesdropping the communication between Alice and Bob will have access to $P$, $Q_a$, and $Q_b$, but cannot reconstruct $k_a$ or $k_b$, as solving the so-called discrete logarithm on elliptic curves does not have an efficient algorithmic solution on (non-quantum) computers.
	In the following we show how this is increases security when using elliptic curves to define the switching function $\sigma_i(\cdot), i \in \{w,q\}$.
	
	\begin{remark}
		The order and cofactor of a generator point $P$ plays a role in how difficult\footnote{A commonly used definition of ``difficulty'' in the IT-based cyber-security literature is that a known algorithm can find a solution in polynomial time.} it is to compute the discrete logarithm problem.
		For cryptographic problems, it is common to select generator points $P$ such that their cofactor $h \leq 4$ \citep{wohlwend2016elliptic}.
		$\hfill\triangleleft$
	\end{remark}
	
	\section{A switching function based on elliptic curves}\label{ch:WM-EC}
	
	Having presented the fundamentals of elliptic curves on finite fields, let us now present the algorithm that defines the switching functions $\sigma$.
	We drop subscripts to improve readibility, which are however to be considered for both watermark generator and remover.
	We then highlight the information to be shared between $\mathcal W$ and $\mathcal Q$ for $\sigma_w = \sigma_q$.
	
	\subsection{Switching function}\label{ch:WM-EC:def}
	Recall from \eqref{eq:WM:sw} that $\sigma(\cdot)$ plays a role in changing the parameters of the watermarking pair, once the triggering function $\tau(y_p(\kappa)) = 1$, for $\kappa\in\mathcal{K}$.
	Here, we take $y_p(\kappa-1)$ as the input to the switching function $\sigma(\cdot)$.
	In the following, we presume that $n_y = 1$, and therefore that $y_p \in \mathbb{R}$. The procedure can be easily extended for $n_y > 1$.
	
	\begin{remark}\label{rem:yP}
		It is necessary to take $y_p(\kappa-1)$ rather than $y_p(\kappa)$ because $y_p(\kappa-1) = y_q(\kappa-1) \in \mathcal I_w \cap \mathcal I_q$, while $y_p(\kappa)\notin\mathcal I_q$, although $y_p(\kappa) \in \mathcal I_w$.
		Indeed, the watermark remover $\mathcal Q$ requires knowledge of the new parameter vector $\theta^+(\kappa)$ to recover the plant's measured output.
		$\hfill\triangleleft$
	\end{remark}
	\begin{remark}
		The switching function depends on $y_p$ to introduce some randomness to the algorithm, thus making it more complex to identify for an attacker.
		Indeed, the use of physical quantities to generate true random numbers (rather than random number generators) is common in the selection of private keys in public key criptography.
		$\hfill\triangleleft$
	\end{remark}

	
	%
	%

	The switching function $\sigma(\cdot)$ is the result of the following:
	\begin{enumerate}[label=\alph*.]
		\item the projection of $y_p(\kappa-1)$ onto $P \in E(\mathbb{F}_s)$, the elliptic curve defined on the field of integers modulo $s$, with $s$ prime;\label{stp:projEC}
		\item the computation of $S = lP$, $S \in E(\mathbb{F}_s)$, for some $l \in \mathbb{Z}_+$; \label{stp:ECmult}
		\item the mapping of $S$ onto $\Theta$, the constrained parameter set guaranteeing that the resulting watermarking pair $\{\mathcal W, \mathcal Q\}$ satisfies conditions in Definition~\ref{def:WM}. \label{stp:projTheta}
	\end{enumerate}
	
	In the first step the measurement output is ``projected'' to a point on the elliptic curve $E(\mathbb{F}_s)$, then used as the generator of a cyclic subgroup.
	We define function $\alpha(\cdot):\mathbb{R}^{n_y} \rightarrow E(\mathbb{F}_s)$ as this function.
	In turn $\alpha(\cdot)$ can be seen as $\alpha(\cdot) = \alpha_2(\alpha_1(\cdot))$, where $\alpha_1(\cdot):\mathbb{R}^{n_y} \rightarrow \mathbb{R}_s \times \mathbb{R}_s$ scales the output, where $\mathbb{R}_s$ is the set of real numbers modulo $s$,
	defining the coordinates of a point $\tilde{P} = (x_{\tilde{P}},y_{\tilde{P}}) \in \mathbb{R}_s \times \mathbb{R}_s$, 
	whilst $\alpha_2: \mathbb{R}_s \times \mathbb{R}_s \rightarrow E(\mathbf{F}_s)$ maps the point $\tilde P$ to $P \in E(\mathbb{F}_s)$.
	Specifically:
	\begin{equation}
	\alpha_1 = (\alpha_{1,x},\alpha_{1,y}),
	\end{equation}
	where $\alpha_{1,x}$ and $\alpha_{1,y}$ may have similar structure, but should have different parameters, such that $x_{\tilde{P}} \neq y_{\tilde{P}}$, in general.
	These functions may be defined in several ways, e.g., that given in Section~\ref{ch:Num}.
	
	Once $\tilde{P}$ is computed, $\alpha_2(\cdot):{\mathbb{F}_s\times\mathbb{F}_s \rightarrow E(\mathbb{F}_s)}$ can be defined as follows:
	\begin{equation}\label{eq:sigma:alph2}
	P = \alpha_2(\tilde P) = 
	\argmin_{\Lambda\in E(\mathbb{F}_s)} \,\|\Lambda-\tilde P\|_2^2,
	\end{equation}
	thus defining a generator point $P\in E(\mathbb{F}_s)$. 
	
	\begin{remark}
		It is important to highlight the importance of scaling the measurement output $y_p$.
		In nominal operations, i.e., whilst in steady state, the output of the plant varies only slightly 
		in a (likely small) neighborhood of some nominal output, $r$, which may be given by a reference to the controller.
		Thus, without scaling by 
		$\alpha_1(\cdot)$, the result of the projection of $y_p$ onto $E(\mathbb{F}_s)$ would always be onto a subset $\mathcal{E} \subset E(\mathbb{F}_s)$, with $|\mathcal E| << |E(\mathbb{F}_s)|$, and possibly $|\mathcal E| = 1$. 
		This is undesirable, as it would imply that $\theta^+ = \theta$.
		Thus, by scaling $y_p$, even small changes in the measured output lead to a difference in $\tilde P$, and therefore to increased variability in $\theta^+$.
		$\hfill\triangleleft$
	\end{remark}

	Once a generator point $P \in E(\mathbb{F}_s)$ is computed, a scalar multiplication is performed to define $S = lP$.
	Note that, given the definition of time-varying generator $P$, it may be that for some values of $P$, $l$ is the order of $P$, and thus that $S = O$, which is undesirable.
	Thus, if $S = O$, some heuristic can be selected, such as $S = P$ or $S = -P = (l-1)P$, to avoid the result $S = O$.
	
	\begin{remark}
		Note that here we assume $l$ to be defined a priori and to be time invariant.
		This is done because, if $l$ were to be time-varying, it would either have to be a function of $y_p(\kappa-1)$, or to satisfy another switching function $l^+ = \pi(l)$.
		While the former is undesirable because it may expose the switching function to attacks capable of reconstructing $y_p$, the latter would require the definition of another switching function.
		$\hfill\triangleleft$
	\end{remark}

	Finally, $\eta(\cdot): E(\mathbb{F}_s) \rightarrow \Theta$ is introduced to map the value of $S$ to an appropriate parameter vector, ensuring that the resulting watermarking pair satisfies Definition~\ref{def:WM}.
	Similarly to 
	$\alpha(\cdot)$, $\eta(\cdot)$ may also be seen as the composition of two functions, $\eta(\cdot) = \eta_2(\eta_1(\cdot))$, where $\eta_1$ maps the points of the elliptic curve into $\mathbb{R}^{n_\theta}$, and $\eta_2$ restricts the novel parameter vector to satisfy $\theta^+ \in \Theta$.
	Once again, in order to ensure that changes in $S$ lead to large differences in $\theta^+$, $\eta_1(\cdot):E(\mathbb{F}_s)\rightarrow \mathbb{R}^{n_\theta}$ should be chosen nonlinear, while the definition of $\eta_2(\cdot):\mathbb{R}^{n_\theta}\rightarrow \Theta$ depends on the class of systems that are used to define $\mathcal W$ and $\mathcal Q$.
	In Section~\ref{ch:WM-EC:stab} we give an example of how to construct $\eta_2(\cdot)$ for the class of finite impulse response (FIR) filters.
	
	Finally, let us comment on what information must be included in $\mathcal I_w$ and $\mathcal I_q$ to ensure that $\sigma_w(\cdot) = \sigma_q(\cdot) = \sigma(\cdot)$, and thus that $\theta_w^+ = \theta_q^+$.
	Recall from Remark~\ref{rem:yP} that $y_p(\kappa-1), \kappa \in \mathcal K$ satisfies
	\begin{equation}
	y_{p}(\kappa -1) \in \mathcal I_w \cap \mathcal I_q,
	\end{equation}
	for all $\kappa \in \mathcal K$, provided $\mathcal K_w = \mathcal K_q$, and $\{\mathcal W, \mathcal Q\}$ satisfy Definition~\ref{def:WM}.
	Thus, it is necessary that
	\begin{equation}
	\{\alpha(\cdot),l,E(\mathbb{F}_s),\eta(\cdot)\} \subset \mathcal I_w \cap \mathcal I_q.
	\end{equation}
	This information is time invariant, to be shared securely at the initial design time of the watermarking system.
	
	\begin{remark}
		Note that in practical applications, attention must be given to the definition of $\alpha(\cdot)$ and $\eta(\cdot)$, to guarantee that numerical issues do not cause errors in the computation of the watermarking parameters, and therefore that $\theta_w^+ \neq \theta_q^+$.
		In-depth analysis of this phenomenon, and methods to solve it are, however, outside of the scope of this paper and will be the subject of future research.
		$\hfill\triangleleft$
	\end{remark}

	\subsection{Security analysis}
	Let us now evaluate the security of the switching function $\sigma(\cdot)$, as defined in Section~\ref{ch:WM-EC:def}.
	Here, we consider ``security'' of $\sigma(\cdot)$ to mean that it is not possible for an attacker to predict, at time $\kappa \in \mathcal K$, what the value of $\theta^+$ is.
	Indeed, even if an attacker had knowledge of $\theta$, and were therefore capable of constructing an undetectable attack, without knowledge of $\theta^+$ its undetectability would be threatened.
	
	In order to formally examine this scenario, let us introduce the set of information known by the attacker, $\mathcal I_a(k), k \geq K_a$.
	This, similarly to the definitions of $\mathcal I_w(k)$ and $\mathcal I_q(k)$, is the set of all signals and parameters that are known to the malicious agent at time $k$. 
	
	To be able to compute $\theta^+$, it is necessary to know $\sigma(\cdot)$, and therefore necessary that:
	\begin{equation}
	\{\alpha(\cdot),\eta(\cdot),l,E(\mathbb{F}_s)\} \subseteq \mathcal I_a.
	\end{equation}
	
	What is particularly interesting, and a direct consequence of using elliptic curves, is that even if all functions were known to the attacker, so long as 
	\begin{equation}
	l \notin \mathcal I_a,
	\end{equation}
	the attacker would not be capable of reconstructing $\theta^+$.
	Indeed, even if an eavesdropping attacker were to be able to estimate $\theta$ and $\theta^+$, after some delay following the switch, and through knowledge of $\alpha(\cdot)$ and $\eta(\cdot)$ were to reconstruct $P$ and $S$, finding $l$ solving $S = lP$ is the solution to the discrete logarithm over elliptic curves, for which there are no known algorithms capable of finding a solution in polynomial time \citep{wohlwend2016elliptic}.

	\begin{algorithm}[t]
		\caption{Switching function $\sigma_w(y_p(\kappa_w-1))$}
		\begin{algorithmic}[1]
			\State \textbf{Input:}  $y_p(\kappa_w-1)$,  $E(\mathbb F_q)$,  $l$,   $\alpha(\cdot)$, $\eta(\cdot)$;
			\State \textbf{Output:} $\theta_w^+$
			\vspace{.1cm}
			\hrule
			\vspace{.1cm}
			\State Compute the generator of the elliptic curve by computing $P = \alpha(y_p(\kappa_w-1))$;
			\State Given $P$, compute $S = lP$;
			\State Define $\theta_w^+(\kappa_w) = \eta(S)$
			\State \textbf{return:} 
			$\theta_w^+(\kappa_w)$
		\end{algorithmic}
		\label{alg:WM:OOG}
	\end{algorithm}

	\subsection{Watermark pair stability: an example with finite impulse response filters}\label{ch:WM-EC:stab}
	Let us now focus on the definition of the parameter set $\Theta$ such that if $\theta_w^+ \in \Theta$, the resulting watermarking pair is guaranteed to satisfy the conditions in Definition~\ref{def:WM}.
	We restrict the set of parameters by giving the watermarking generator and remover some \textit{structure}, namely, for the purpose of this paper, we suppose that $\mathcal W$ is composed of $n_y$ \textit{parallel} FIR filters of order $n_h$; thus, $n_\theta = n_y \cdot n_h$.
	For the sake of maintaining notation streamlined, and without loss of generality, for the remainder of this subsection we suppose that $n_y = 1$.
	The output of $\mathcal W$ in \eqref{eq:WM:W} is thus:
	\begin{equation}\label{eq:WM:W:fir}
	y_w(k) = \sum_{h = 0}^{n_h}b_{h} y_{p,i}(k-i),
	\end{equation}
	with $b_h \in \mathbb{R}, \forall h \in \{0,1,\dots,n_h\}$. This formulation guarantees that $\mathcal W$ is stable, with $n_h$ poles at the origin.
	Finding $\Theta \subseteq \mathbb{R}^{n_h}$ is equivalent to finding the set of parameters $b_h$ for which $\mathcal W$ is invertible, with stable inverse.

	\begin{theorem}\label{th:WM:stab}
		Suppose that $\mathcal W$ is an LTI system with dynamics defined by the FIR filter in \eqref{eq:WM:W:fir}.
		Thus, if the parameters are such that
		\begin{subequations}\label{eq:WM:stab}
			\begin{gather}
			b_0 \neq 0 \label{eq:WM:stab:b0}\\
			|b_1| < 1\label{eq:WM:stab:b1}\\
			\sum_{i = 2}^{n_h} \left|\frac{b_i}{b_0}\right| < 1 - |b_1|\label{eq:WM:stab:rad}
			\end{gather}
		\end{subequations}
		hold, the resulting watermarking pair $\{\mathcal W,\mathcal Q\}$ is guaranteed to satisfy Definition~\ref{def:WM}, with $\theta_w = \theta_q = \mathrm{col}_i [b_i]$.
		$\hfill\square$
	\end{theorem}
	\begin{proof}
		We start by noting that if $\mathcal W$ is defined by \eqref{eq:WM:W:fir}, then it is by definition stable.
		Indeed, FIR filters have all poles at the origin.
		Furthermore, if \eqref{eq:WM:stab:b0} is satisfied, $\mathcal W$ admits an inverse.
		We therefore must prove that the watermark remover $\mathcal Q = \mathcal W^{-1}$ is stable.	
		The dynamics of $y_w$ defined in \eqref{eq:WM:W:fir} can be written in state-space form \eqref{eq:WM:W} with the following matrices:
		\begin{equation}
		\begin{array}{cccc}
		A_w = \matrices{0 &0 &\dots &0 \\
			1 &0 &\dots &0\\
			\vdots &\ddots &\ddots &\vdots\\
			0 &\dots &1 &0
		} &
		B_w = \matrices{1\\0\\\vdots\\0}
		& C_w = \matrices{b_1\\b_2\\\vdots\\b_h}^\top
		&D_w = b_0.
		\end{array}
		\end{equation}
		Recalling the definition of $\mathcal Q$ as the inverse of $\mathcal W$, given the definition of the inverse of a system \citep[Lemma 3.15]{zhou1996robust}, we write the matrices defining the system dynamics \eqref{eq:WM:Q} as:
		\begin{equation}
		\begin{array}{cc}
		\vspace{.4cm}
		A_q = \frac{1}{b_0}\matrices{-b_1 &-b_2 &\dots &-b_h\\
			b_0 & 0 &\dots &0\\
			\vdots &\ddots &\ddots &\vdots\\
			0 &\dots &b_0 &0} &
		B_q = -b_0^{-1}B_w \\
		C_q = b_0^{-1}C_w  &D_q = b_0^{-1}.
		\end{array}
		\end{equation}
		
		Thus, by Ger{\v{s}}gorin's circle theorem \citep{horn1985matrix}, it is sufficient that conditions \eqref{eq:WM:stab} hold for $\mathcal Q$ to be stable.
		Hence, $\{\mathcal W,\mathcal Q\}$ satisfy Definition~\ref{def:WM}.
		$\hfill\blacksquare$
	\end{proof}
	
	Conditions \eqref{eq:WM:stab} in Theorem~\ref{th:WM:stab} define $\Theta$ for the class of FIR filters.
	We can therefore now define an example for $\eta(\cdot)$.
	Take $\eta_1(\cdot):E(\mathbb{F}_s) \rightarrow \mathbb{R}^{n_h}$ to be any vector of $n_h$ nonlinear functions, and define:
	\begin{equation}
	b^- = \eta_1(S).
	\end{equation}
	Define each element of $b^- \in \mathbb{R}^{n_h}$ as $b_i^-$.
	To ensure that $b = \eta_2(b^-)$ satisfies \eqref{eq:WM:stab} in Theorem~\ref{th:WM:stab}:
	\begin{enumerate}[label=\roman*.]
		\item condition \eqref{eq:WM:stab:b0} is satisfied if \begin{equation}\label{eq:sigma:eta2:1}
		\eta_{2,0}(S) \neq 0, \forall S \in E(\mathbb{F}_s);
		\end{equation}
		\item condition \eqref{eq:WM:stab:b1} holds if 
		\begin{equation}
		\eta_{2,1}(b_1^-) = \frac{b_1^-}{|b_1^-|+\eta},
		\end{equation}
		with $\eta \in \mathbb{R}_+$ arbitrary;
		\item condition \eqref{eq:WM:stab:rad} is satisfied for
		\begin{equation}\label{eq:sigma:eta2:3}
		\eta_{2,i}(b^-) = \xi(b^-) b_{i}^-,
		\end{equation}
		where $\xi(b^-)$ is an auxiliary variable:
		\begin{equation}
		\xi(b^-) = \frac{1-|b_1|-\varepsilon}{\sum_{h = 2}^{n_h}\left|\frac{b_h^-}{b_0}\right|},
		\end{equation}
		with $\varepsilon \in (0,1-|b_1|)$ arbitrary.
	\end{enumerate}

	\section{Numerical results}\label{ch:Num}
	Having presented the switching function, and given an overview of its properties, let us now focus on considerations for the practical implementation of $\sigma(\cdot)$.
	We start by discussing the consequences of the definition of both the elliptic curve and of the functions $\alpha(\cdot)$ and $\eta(\cdot)$.
	Following this, we highlight the effect that different operating points have on the \textit{sensitivity} of $\sigma(\cdot)$ to changes in its inputs.
	
\begin{figure}[t]
		\centering
		\includegraphics[width=0.85\linewidth]{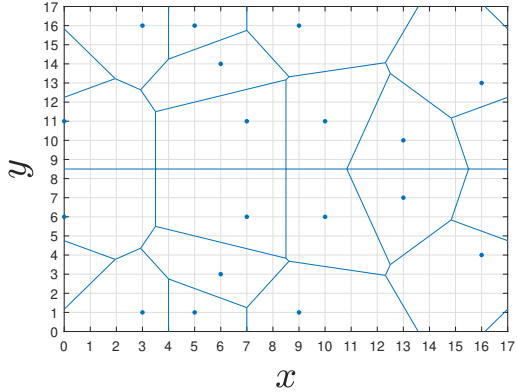}
		\caption{
			The points define the elliptic curve in $\mathbb{F}_{17}$, with $a = 2$ and $b = 2$.
			Superimposed is the Voronoi partition with points of the elliptic curve as the seeds of all cells.
		}
		\label{fig:EC_fin}
	\end{figure}

	The elliptic curve we choose for this example is that defined on $\mathbb{F}_{17}$, shown in Figure~\ref{fig:EC_fin}.
	This is an elliptic curve, of order $|E(\mathbb{F}_{17})| = 19$, on which the solution to the discrete logarithm problem does not require a lot of computation; indeed, in cryptographic settings, \cite{fips2013digital} recommends using elliptic curves defined on fields where the prime integer used for the modulo operation is at least 192 bits long.
	However, its structure makes it suitable for illustrating some fundamental characteristics of $\sigma(\cdot)$:	indeed,
	because its order is a prime number, the order of each of its points $P \in E(\mathbb{F}_{17})$ is also $19$, as the coefficient $h$ of $P$, defined in Definition~\ref{def:EC:cof}, is always an integer \citep{birkhoff2017survey}.
	Furthermore, by selecting $l$ such that $l \mod 19 \neq 0$, it is possible to guarantee that $S = lP \neq O$, the point at infinity, for all $P \in E(\mathbb{F}_{17})$. 
	Moreover, the order of $E(\mathbb{F}_{17})$ also determines the number of parameters $\theta_w = \theta_q \in \Theta$ that define $\{\mathcal W, \mathcal Q\}$.
	It is worth noting that the points $P \in E(\mathbb{F}_{17})$ do not partition the space uniformly. 
	To show this, in Figure~\ref{fig:EC_fin} we include the Voronoi diagram generated by taking the points in $E(\mathbb{F}_{17})$ as seeds: there are points (and thus parameters) that have a higher likelihood of being selected by a given $\gamma\in\mathbb{R}^{n_y}$.

	We can now focus our attention on the definition of $\alpha(\cdot)$ and $\eta(\cdot)$, supposing $n_y = 1$.
	As discussed previously, both of these can be seen as the combination of a scaling and a projection function.
	While the latter have been discussed in Section~\ref{ch:WM-EC:def}, a possible definition of the former is:
	\begin{align}
	&\alpha_{1,i}(\gamma) = a_{i,0} \mathrm{atan}(a_{i,1} \gamma) + \sum_{j = 2}^{n_\alpha} a_{i,j} |\gamma|^j\\
	&\eta_1(H) = \sum_{j = 0}^{n_\eta} b_j \|H\|_2^j
	\end{align}
	with $H \in E(\mathbb{F}_{17})$, and where $a_{i,j}, i \in \{x,y\}, j \in \{0,1,\dots,n_\alpha\}$ and $b_j, j \in \{0,1,\dots, n_\eta\}$ are time-invariant parameters, shared between $\mathcal W$ and $\mathcal Q$; moreover, recall from Section~\ref{ch:WM-EC:def} that $\alpha_1(\cdot) \doteq (\alpha_{1,x}(\cdot),\alpha_{1,y}(\cdot))$.
	
\begin{figure}[t]
		\centering
		\includegraphics[width=0.85\linewidth]{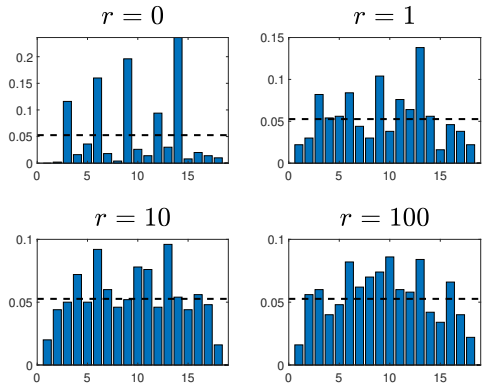}
		\caption{
			Histograms giving, for each point in the elliptic curve $E(\mathbb{F}_{17})$, the relative frequency of a point being reached from $\alpha_1(y)$, for $y = r+v$, with $r \in \{0,1,10,100\}$, and $v$ a sample of a uniform distribution with limits $\pm 0.05$.
			For each reference point, $500$ realizations of $v$ are taken.
			The horizontal black-dashed lines are set at the height of a uniform variability over all points.
		}
		\label{fig:EC_var_hist}
	\end{figure}

	Finally, we are interested in examining the sensitivity of the output of $\sigma(\cdot)$ with respect to (small) variations of the plant measurement outputs $y_p$.
	Results are presented in 
	Figure~\ref{fig:EC_var_hist}.
	We consider the measurement of the plant $y_p = r + \epsilon$, where $r \in \mathbb{R}^{n_y}$ is an output reference for $\mathcal P$, and $\epsilon$ is the tracking error.
	Specifically, we suppose $r \in \{0,1,10,100\}$, and for each we consider $500$ realizations of $v$, each taken from a uniform distribution with limits $[-0.05,0.05]$.
	In Figure~\ref{fig:EC_var_hist} we show how frequently different points $P \in E(\mathbb{F}_{17})$ are reached for changes in $y_p$, relative to the total number of values of $v$ taken.
	We see that, irrespective of the operating point, it is possible to reach all cells, which implies that for all operating conditions all $19$ values that $\theta^+$ can take are reachable.
	However, 
	there are large differences between the behaviors relative to the reference points themselves.
	Indeed, for $r = 0$, it is significantly more likely that a subset $\mathcal E$ of all points in the elliptic curve is reached, with $|\mathcal E| = 5$.
	On the other hand, for $r = 10$ the probability of reaching a given point in the elliptic curve is closer to being uniform across all points.
	In this case, $\sigma(\cdot)$ is more sensitive to small changes in the input.
	This gives us important insight for the practical implementation of $\sigma$: indeed, $\alpha_1(\cdot)$, which is the main ``driver'' of this sensitivity, is to be appropriately tuned for those points that are to be tracked by the plant, to ensure there is sufficient sensitivity of the switching function $\sigma(\cdot)$.



	\section{Conclusion}
	In this work, we have presented a method to define the switching function for switching multiplicative watermarking based on elliptic curves, guaranteeing that . We show how, even for attackers with large amounts of information, the switching signal may remain secure.
	
	In future work, we are interested in testing the real-world applicability of this scheme, testing it on real world hardware, considering computational and numerical problems.

	\bibliographystyle{IEEEtran}
	\bibliography{cryptoWM}
	
\end{document}